\documentclass[twocolumn,showpacs,pra,aps,eqsecnum]{revtex4}

\usepackage{amsmath}
\usepackage{amssymb}
\usepackage{graphicx}
\usepackage{upgreek}
\usepackage{bm}

\newcommand{\Tr}{\mathrm{tr}}

\newcommand{\rmd}{\mathrm{d}}
\newcommand{\rmi}{\mathrm{i}}
\newcommand{\rme}{\mathrm{e}}

\newcommand{\dyad}[1]{\mathbf{#1}}
\newcommand{\dGamma}{\Gamma}

\newcommand{\vc}[1]{\mathbf{#1}}

\newcommand{\jsum}{\sum_{j=0}^\infty \!{}^{{}^\prime}}

\newcommand{\kB}{k_\mathrm{B}}
\newcommand{\ep}{\varepsilon}
\newcommand{\PC}{\mathrm{PC}}

\newcommand{\Unr}{U^\mathrm{nr}}
\newcommand{\Ur}{U^\mathrm{r}}

\newcommand{\order}{\mathcal{O}}

\newcommand{\rs}{r^\mathrm{TE}_l}
\newcommand{\rp}{r^\mathrm{TM}_l}
\newcommand{\rsPC}{r^{\mathrm{TE},\PC}_l}
\newcommand{\rpPC}{r^{\mathrm{TM},\PC}_l}
\newcommand{\hl}{h_l^{(1)}}
\newcommand{\thl}{\tilde{h}_l^{(1)\prime}}
\newcommand{\tjl}{\tilde{\jmath}'_l}
\newcommand{\sqe}{\sqrt{\varepsilon}}

\newcommand{\half}{{\textstyle\frac1{2}}}

\newcommand{\be}{\begin{equation}}
\newcommand{\ee}{\end{equation}}

\newcommand{\re}{\mathrm{Re}}
\newcommand{\im}{\mathrm{Im}}

\begin{document}

\title{Casimir-Polder energy level shifts of an out-of-equilibrium
particle near a microsphere}

\author{Simen {\AA}. Ellingsen}
\affiliation{Department of Energy and Process Engineering, Norwegian
University of Science and Technology, N-7491 Trondheim, Norway}
\author{Stefan Yoshi Buhmann}
\author{Stefan Scheel}
\affiliation{Quantum Optics and Laser Science, Blackett Laboratory,
Imperial College London, Prince Consort Road, London SW7 2BW, United
Kingdom}

\date{\today}

\begin{abstract}
Rydberg atoms and beams of ultracold polar molecules have become
highly useful experimental tools in recent years. There is therefore a
need for accessible calculations of interaction potentials between
such particles and nearby surfaces and structures, bearing in mind
that the particles are far out of thermal equilibrium with their
environment and that their interaction is predominantly non-retarded.
Based on a new perturbative expansion with respect to the inverse
speed of light and the inverse conductivity, we derive a simple,
closed-form expression for the interaction potential (i.e., the
particle energy level shifts) of a particle and a metallic sphere
that is is accurate at better than 1\% level for typical experimental
set-ups at room temperature and above, and off by no more than a few
percent at any temperature including zero. Our result illuminates the
influence of retardation and imperfect conductivity and the interplay
of these effects with geometry.
The method developed for the present study may be applied to other,
more complex geometries.
\end{abstract}

\maketitle

\section{Introduction}

Recent times have witnessed a blossoming of experimental set-ups
in which the detailed interaction of particles
with nearby surfaces, in particular the Casimir--Polder (retarded van
der Waals) \cite{casimir48} interaction, is important. Some such
systems are typically far out of thermal equilibrium,
such as Bose-Einstein condensates in magnetic traps close to surfaces,
beams of cold polar molecules and Rydberg atoms.
We shall focus on the two latter categories herein. For example, the
interaction between Rydberg atoms \cite{gallagher94} and surfaces are
essential to the understanding of the behavior of Rydberg atoms in
vapour cells \cite{kubler10} and near atom chips \cite{tauschinsky10},
systems which have already been investigated in several experiments.
Various suggested mechanisms for quantum information processing also
involve Rydberg atoms close to metallic
surfaces \cite{sorensen04,hyafil04,petrosyan09}. Moreover, beams of
cold polar molecules have already been put to use in a range of
experimental applications as reviewed in
Refs~\cite{vandemeerakker08,carr09,bell09}. For instance, trapping of
cold CO molecules near atom chips using electric traps has recently
been realized \cite{meek09}.

We derive herein a simple closed form expression for the CP
interaction between a particle and a sphere valid both for Rydberg
atoms and cold molecules. Second only to the plane surface, the
spherical geometry is arguably the most generically useful to consider
in all microscopic applications. The microsphere is the standard
vehicle in the rapidly progressing field of micromanipulation and
photonics, a field closely bordering on atomic physics where CP forces
are of importance. Using laser beams, microspheres can be trapped 
and pushed \cite{ashkin70} and perhaps even pulled \cite{chen11} for
detailed manipulation, and are readily transported along optical
fibres via the evanescent field \cite{kawata92}. Microsphere optical
resonators with extremely high Q-factors have been built,
and are useful e.g.\ for low threshold lasing \cite{min03}. Our closed
form expression, not involving the typical lengthy sums of Mie
scattering coefficients, is immediately useful for direct insertion
into numerical simulations of microsystems, as well as analysis of
experimental data.

Rydberg atoms and cold (ground state) polar molecules share two traits
that set them sharply apart from the ground state or thermalized atoms
which have typically been considered in the van der Waals and
Casimir--Polder (CP) literature. Firstly, they are both far out of
thermal equilibrium with their thermal environment. Rydberg atoms have
been excited to a high principal quantum number, far from an atom's
thermalized state, which is almost identical to its ground state since
excitation energies are large compared to $\kB T$ (room temperature
assumed). The excitation energies of rovibriational states of
molecules, in contrast, are small compared to $\kB T$, so a
thermalized molecule significantly occupies a number of its energy
eigenstates. Thus also ground state molecules at room temperature are 
far from thermal equilibrium. Systems with magnetic transitions
exhibit similar properties \cite{Haakh}.

Secondly, and for the same reason, the Casimir-Polder interaction
between these particles and nearby surfaces is predominantly
non-retarded, and retardation corrections due to the finite speed of
light enter only as a correction. Typically, the retarded interaction
regime stretches for tens and hundreds of micrometers for polar
molecules and Rydberg atoms, respectively \cite{buhmann08b,crosse10},
thus including the separations normally encountered in experiments and
applications.

Whereas thermal non-equilibrium initially complicates theoretical
treatment, the non-retardedness of the interaction introduces a
significant simplification, allowing almost surprisingly simple
results to be achieved. The general theory for Casimir-Polder
interactions of a particle in an arbitrary superposition of
eigenstates was recently derived by some of us \cite{buhmann08}, and
has since been applied to planar geometries for Rydberg atoms and
molecules \cite{crosse10,ellingsen09}. Strikingly, it was found that
for non-retarded interaction with a flat metallic surface the
interaction potential is virtually independent of temperature
\cite{ellingsen10}, a result that can be extended to arbitrary
geometries \cite{ellingsen11}.
For a full theoretical background of different non-equilibrium CP
scenarios the reader may additionally refer to
Refs.~\cite{nakajima97,wu00,gorza06,antezza08,sherkunov09,%
buhmann07,scheel08}. The theories for different non-equilibrium
situations, albeit apparently disparate, may be shown
to concord as they should \cite{ellingsen10Q}.

We consider the general situation of a particle whose eigenstates 
are $|n\rangle$. It was shown in Ref.~\cite{buhmann08} that for a
particle prepared in an arbitrary superposition of eigenstates
$|\phi\rangle=\sum_n p_n |n\rangle$ with occupation probabilities
$p_n$, the Casimir-Polder potential may be written as a sum over
transitions between pairs of eigenstates according to
\be
U_\phi
 = \sum_n p_n U_n
\ee
with
\be\label{Un}
  U_n = \sum_k U_{nk}
\ee
where the sum runs over all other eigenstates $|k\rangle$ to which
there is an allowed dipole transition. 

Crucial to the understanding of the CP interactions of both 
cold molecules \cite{ellingsen09} and Rydberg atoms \cite{crosse10}
is the realization that
only a few transitions turn out to give significant contributions. To
wit, the important transitions were found to be those corresponding to
the smallest difference in eigenenergy $\Delta E_{kn} = E_k-E_n$,
i.e.\ the smallest transition frequency $\omega_{kn}=\Delta
E_{kn}/\hbar$ or correspondingly the longest transition wavelength
$\lambda_{kn}=2\pi c/\omega_{kn}$.
For example, a Rydberg atom near a half-space prepared in an $s$-state
of principal quantum number $n$, obtains significant contributions
from transitions to the few different $p$-states of principal
quantum numbers $n$ and $n-1$ \cite{crosse10}; for ground state LiH
molecules the only significant transition was to the lowest rotational
state, whereas for YbF also the first vibrational state was required
\cite{ellingsen09}. Transitions with larger $\Delta E_{kn}$ could be
ignored to a good approximation.

Because of this fact the typical values of $\lambda_{kn}$ for cold
polar molecules and Rydberg atoms alike are usually much larger than
the typical distance $z$ from the particle to a nearby body in
experiments involving surfaces. In other words, for the dominating
transitions $|n\rangle\to |k\rangle$,
\be\label{nonret}
  \frac{z}{\lambda_{kn}} = \frac{\omega_{kn}z}{2\pi c} 
  \ll 1,
\ee
hence the interaction is essentially \emph{non-retarded}.

Recently we found that in the non-retarded regime the CP
interaction near a metallic half-space is virtually temperature
independent \cite{ellingsen10}. The thermal CP potential then agrees
with its zero-temperature counterpart for all temperatures. This was
later shown to be a reasonable approximation for bodies of arbitrary
shape \cite{ellingsen11}. Temperature-dependent corrections to the
zero-temperature potential were identified to stem from retardation
and imperfect conductivity. The magnitude of the latter corrections
were found to strongly depend on the body shape and curvature,
demonstrating that geometry and temperature are closely intertwined
\cite{weber10}. It is therefore necessary in practice to study
different geometries individually.

The case of an atom interacting with a metal sphere to be studied
is a prototype of a body with a curved surface. Various embodiments of
the particle--sphere interactions at zero temperature have been
treated by a number of authors
\cite{ferrell80,marvin82,girard89,jhe95,klimov96,buhmann04,taddei10,
sambale10}. 
In most of these works, the CP potential is obtained from a numerical
computation which suffers from poor convergence at small curvatures.
In contrast, we will derive an approximate analytical result based on
a perturbative expansion that is readily accessible while illuminating
the impact of retardation and imperfect reflection as well as the
interplay of these factors with geometry: We calculate the CP
potential contribution from transition $|n\rangle \to |k\rangle$ under
the relevant assumption of non-retarded interaction,
Eq.~\eqref{nonret}, which we quantify by a retardation parameter
\be\label{xdef}
  x = \frac{r\omega_{kn}}{c} \ll 1
\ee
where $r$ is distance from the particle to the sphere's center. We
quote here the final result, to be derived below, for the thermal
Casimir--Polder potential of an atom at distance $r$ from the center
of a metal sphere of radius $R$:
\begin{widetext}
\begin{align}\label{finalresult}
  U_{nk}(r)
 =& -\frac{|\vc{d}_{kn}|^2}{24\pi \varepsilon_0 r^3}\frac{\phi^3
 (6-3\phi^2 + \phi^4)}{(1-\phi^2)^3}
  + \frac{|\vc{d}_{kn}|^2 }{24\pi \varepsilon_0 r^3}
 \left(\frac{\kB T}{\hbar\omega_{kn}}-\frac1{2}\right)\Big\{ x^2\bigl[
 3(1+3\phi^4)\mathrm{artanh}\phi-\phi(3-\phi^2)\notag \\
  &+2\phi^3\log(1-\phi^2)\bigr]
+ 2x\phi^2\re\Bigl\{\frac{i}{\sqrt{\varepsilon(\omega)}}\Bigr\}
\Bigl[\frac{3+7\phi^2 -
4\phi^4}{(1-\phi^2)^2}-\log(1-\phi^2)\Bigr]+...\Bigr\} +
\order(T^{-1})
\end{align}
\end{widetext}
where we have introduced the dimensionless geometry parameter
\be \label{phidef}
  \phi = R/r.
\ee
The expression~\eqref{finalresult} 
is remarkably simple compared to a numerical evaluation of the 
starting equations, involving infinite sums over Mie scattering 
coefficients. It
holds when $x\ll 1$, but
$\im\{\sqe\} x\phi=\im\{\sqe\}\omega_{kn}R/c$ still significantly
exceeds unity, which is the case for good conductors in combination
with typical values of $x$ for the systems under consideration.
The dots indicate higher-order contributions in the small parameters
$x$ and $1/(x\sqe)$. For definitions of the various quantities in
Eq.~\eqref{finalresult}, see Sec.~\ref{sec:generalformalism} below.

In the following we derive the CP potential for a particle near a
metallic sphere including the leading correction for small $x$
(retardation correction) and $1/(x\sqe)$ (imperfect reflectivity
correction), starting from the general CP theory for particles out of
thermal equilibrium, which is outlined in Section
\ref{sec:generalformalism}. 
Our method may in principle be employed for any geometry to derive
perturbative temperature corrections such as that presented herein.
Explicit corrections for the particle--sphere configuration are
derived in Section \ref{sec:particlesphere} and analysed in Section
\ref{sec:discussion}.

%%%%%%%%%%%%%%%%%%%%%%%%%%%%%%%%%%%%%
%%%%%%%%%%% S E C T I O N %%%%%%%%%%%
%%%%%%%%%%%%%%%%%%%%%%%%%%%%%%%%%%%%%
\section{General formalism}\label{sec:generalformalism}

The general formalism for the temperature-dependent CP force on a
particle in an energy eigenstate is found in Ref.~\cite{buhmann08}.
Here we shall restrict our attention to the special case of an
isotropic particle.
As explained above, the potential on a Rydberg atom or cold molecule
alike takes form of a sum over just a few contributions from pairs 
of eigenstates which all have transition wavelengths in the same order 
of magnitude. It is sufficient therefore to consider a single
transition $|n\rangle\to |k\rangle$. Our considerations for one such 
transition will therefore hold for all relevant transitions, and all 
that is required in order to return to the full description of these 
particles is to sum the final result over the relevant transitions 
according to Eq.~\eqref{Un}. The generalisation to anisotropic
particles is straightforward (cf.~Refs~\cite{buhmann08,ellingsen11}).

This section reviews the general framework for treating the small
temperature correction in the non-resonant regime for arbitrary
geometries, which we apply below to a metal sphere. In the notation of
Ref.~\cite{ellingsen11}, the CP potential of an isotropic non-magnetic
particle in state $|n\rangle$ due to a possible transition to state
$|k\rangle$ splits naturally into a non-resonant (nr) and a resonant
(r) part \cite{buhmann08}
\be
  U_{nk}(\vc{r}) = \Unr_{nk}(\vc{r})+\Ur_{nk}(\vc{r}),
\ee
where the two parts are given by
\begin{subequations}
\begin{align}
  \Unr_{nk}(\vc{r})
 =& -\frac{2\kB T|\vc{d}_{nk}|^2\omega_{kn}}{3\hbar\varepsilon_0}\jsum
\frac{\dGamma_{\rmi\xi_j}(\vc{r})}{\omega_{kn}^2+\xi_j^2};
\label{eq:nonres}\\
 \Ur_{nk}(\vc{r})=&\frac{|\vc{d}_{kn}|^2}{3\varepsilon_0}\,
n(\omega_{kn})\re\,\dGamma_{\omega_{kn}}(\vc{r})
\label{eq:res}.
\end{align}
\end{subequations}
Here, $\vc{d}_{kn} = \langle k|\vc{d}|n\rangle$ is the transition
dipole matrix element and 
\be
  \xi_j =
2\pi j\kB T/\hbar
\ee
are the Matsubara frequencies. The photon number at frequency $\omega$
and temperature $T$ is given by the Bose-Einstein distribution,
\be
  n(\omega) = \frac{1}{\exp(\hbar\omega/\kB T)-1} =
-[n(-\omega)+1].
\ee
The function
\be
  \dGamma_\omega(\mathbf{r})
  \equiv
\frac{\omega^2}{c^2}\lim_{\mathbf{r}'\to\mathbf{r}}
\Tr\dyad{G}^{(1)}(\mathbf{r},\mathbf{r}',\omega)
\label{eq:gamma}
\ee
is given in terms of the scattering part $\dyad{G}^{(1)}$ of the
total dyadic Green's function satisfying
\be
  \left[\nabla\times\nabla\times
 ~ -
\frac{\omega^2}{c^2}
\varepsilon(\mathbf{r},\omega)
\right]\dyad{G}(\mathbf{r},\mathbf{r}',\omega) =
\delta(\mathbf{r}-\mathbf{r}'){\boldsymbol 1}.
\ee
($\boldsymbol 1$: unit tensor). The relative permittivity
$\varepsilon(\mathbf{r},\omega)$ of the present bodies is isotropic
and we have assumed the bodies to be non-magnetic. Due to causality,
$\dGamma_{\rmi \xi_j}$ is real, being a generalized susceptibility
evaluated at imaginary frequency, so in particular $\dGamma_{\rmi
0}=\dGamma_{0}$ is real.

In the following, we consider a single transition $|n\rangle\to
|k\rangle$ and simplify our notation according to
\[
  |\vc{d}_{kn}|^2\to |\vc{d}|^2;~~
  \omega_{kn}\to \omega; ~~
  U_{nk}(\vc{r}) \to U(\vc{r}).
\]
Note that $\omega$ can be either positive or negative depending
on whether the transition is upwards or downwards.

Knowing that the potential is largely temperature independent
through the non-retarded region, consider for now the regime in which the
linear $T$-corrections becomes important, i.e., the spectroscopic 
high-temperature regime,
\[
  \kB T\gg \hbar \omega.
\]
Here, the contribution of the lowest Matsubara frequency ($j=0$)
dominates in Eq.~\eqref{eq:nonres}, so
\be
  \Unr(\vc{r}) =
  -\frac{|\vc{d}|^2}{3\varepsilon_0}\frac{\kB T}{\hbar\omega}
  \dGamma_0(\vc{r}) + \order(T^{-1}).
\ee
The photon number in this regime is
\be
  n(\omega) = \frac{\kB T}{\hbar\omega}-\frac1{2} + \order(T^{-1}),
\ee
so the resonant potential~\eqref{eq:res} reads
\begin{equation}
  \Ur(\vc{r})
=\frac{|\vc{d}|^2}{3\varepsilon_0}
\left(\frac{\kB T}{\hbar\omega}-\frac1{2}\right)
\re\dGamma_\omega(\vc{r})\label{eq:res2}.
\end{equation}
Combining these results, we find for the full potential in the
spectroscopic high-temperature regime that
\begin{align}\label{Tcorr}
  U(\vc{r}) 
  =& -\frac{|\vc{d}|^2}{6\varepsilon_0}\dGamma_0(\vc{r})
  + \frac{|\vc{d}|^2}{3\varepsilon_0}
  \left(\frac{\kB T}{\hbar\omega}-\frac{1}{2}\right)
  \re\Delta\dGamma_\omega(\vc{r}) \notag \\
  &+\order(T^{-1})
\end{align}
where $\Delta\dGamma_\omega = \dGamma_\omega-\dGamma_0$.

For comparison, in the zero-temperature limit, in which the Matsubara
sum becomes an integral according to standard procedures (e.g.\ the
Euler-Maclaurin formula), one obtains
\begin{align}
  U(\vc{r})\bigr|_{T=0}
  =&  -\frac{|\vc{d}|^2\omega}{3\pi\varepsilon_0}
  \int_0^\infty\rmd \xi \frac{\Tr\dGamma_{\rmi \xi}^{(1)}(\vc{r})}
 {\omega^2+\xi^2}\notag \\
&-\frac{|\vc{d}|^2}{3\varepsilon_0}\,
  \Theta(-\omega)\re\,
 \dGamma_{\omega}(\vc{r})\label{Tzero}
\end{align}
with $\Theta(x)$ denoting the unit step function.

In the nonretarded and perfect-conductor limits, we have
$\dGamma_{\rmi \xi}(\vc{r})\simeq\re\dGamma_\omega(\vc{r})%
\simeq\dGamma_0(\vc{r})$ \cite{ellingsen10} which implies
$\re\Delta\dGamma_\omega(\vc{r})\simeq 0$. In this case, both
Eqs.~\eqref{Tcorr} and \eqref{Tzero} reduce to
\begin{equation}\label{Tinv}
  U_0(\vc{r})
  = -\frac{|\vc{d}|^2}{6\varepsilon_0}
\dGamma_0(\vc{r})
\end{equation}
and the CP potential is independent of temperature throughout.

In this article, we are interested in the corrections to the
temperature-independent result~\eqref{Tinv} that arise due to small
violations of the non-retarded limit and perfect reflectivity. As seen
from Eq.~\eqref{Tcorr}, they are governed by
$\re\Delta\dGamma_\omega(\vc{r})$. When all present macroscopic bodies
are perfectly conducting (PC), then $\dGamma_\omega$ satisfies
\cite{ellingsen11}
\be\label{retCorr}
  \dGamma_\omega^\PC = \dGamma^\PC_0 +
\frac{\omega^2}{2c^2}\frac{\rmd^2 \dGamma^\PC_\omega}{\rmd \omega^2}
\Bigr|_{\omega=0} + ...~; ~~ \omega\to 0,
\ee
so that
\be\label{retCorr2}
  \Delta\dGamma_\omega^\PC \approx\frac{\omega^2}{2c^2}\frac{\rmd^2
\dGamma^\PC_\omega}{\rmd \omega^2}
\Bigr|_{\omega=0}; ~~ \omega\to 0,
\ee
is quadratic in $\omega$. This correction accounts for the fact that
electromagnetic interactions are transmitted at the finite speed of
light; we will refer to it as the retardation correction in the
following.

For an imperfect conductor, the corrections to the Green's function
for small frequencies includes a second correction due to the
frequency-dependence of the reflectivity of the bodies. We
write
\be\label{DGamma}
 \Delta\dGamma_\omega
 = \Delta\dGamma_\omega^\text{ret.} +
 \Delta\dGamma_\omega^\text{refl.}
\ee
due to retardation and reflectivity, respectively.
When treating $\omega$ and $\varepsilon(\omega)$ as independent
variables, the retardation correction
$\Delta\dGamma_\omega^\text{ret.}$ is the leading-order term in
$1/\sqrt{\varepsilon(\omega)}$ and next-to-leading in $\omega$;
whereas the reflectivity correction
$\Delta\dGamma_\omega^\text{refl.}$ is the contribution sub-leading in
$1/\sqrt{\varepsilon(\omega)}$ and leading in $\omega$. Note that the
perfect-conductor limit $|\varepsilon(\omega)|\to\infty$ does not
commute with the nonretarded limit $\omega\to 0$ in this case. The
incompatibility of the two limits was first pointed out in
Ref.~\cite{babikker76} and is at the heart of the debate over the
temperature correction to the Casimir effect \cite{brevik06}.
For a metal body at typical frequencies and distances, the
perfect-conductor limit has to be performed before the
nonretarded limit, see Sect.~\ref{sec:particlesphere} below.

With the leading corrections to the Green's function being given by
Eq.~\eqref{DGamma}, the thermal CP potential~\eqref{Tinv} can be
given as
\begin{align}\label{Tcorr2}
  U(\vc{r})
  =U_0(\vc{r})+\Delta U_\text{ret.}(\vc{r})
  +\Delta U_\text{refl.}(\vc{r})
  +\order(T^{-1})
\end{align}
with
\begin{equation}\label{DUi}
\Delta U_i(\vc{r})
=\frac{|\vc{d}|^2}{3\varepsilon_0}
\left(\frac{\kB T}{\hbar\omega}-\frac{1}{2}\right)
\re\Delta\dGamma_\omega^i(\vc{r}),
\end{equation}
$i=\text{ret.},\text{refl.}$ The relative corrections due to
retardation and reflection read
\begin{equation}\label{DUret}
\frac{\Delta U_i(\vc{r})}{U_0(\vc{r})}
  = -2\left(\frac{\kB T}{\hbar\omega}-\frac{1}{2}\right)
 \frac{\re\Delta\dGamma_\omega^i(\vc{r})} {\dGamma_0(\vc{r})}\,.
\end{equation}
Note that $\re\Delta\dGamma_\omega(\vc{r})$ is an even function of
$\omega$, as follows directly from definition~\eqref{eq:gamma}
together with the Schwarz reflection principle
$\dyad{G}(\mathbf{r},\mathbf{r}';-\omega)
=\dyad{G}^\ast(\mathbf{r},\mathbf{r}';\omega)$. As a consequence, the
leading temperature corrections in the high-temperature limit, being
proportional to $\re\Delta\dGamma_\omega(\vc{r})/\omega$, change
sign when comparing downward and upward transitions.

%%%%%%%%%%%%%%%%%%%%%%%%%%%%%%%%%%%%%
%%%%%%%%%%% S E C T I O N %%%%%%%%%%%
%%%%%%%%%%%%%%%%%%%%%%%%%%%%%%%%%%%%%
\section{Casimir--Polder potential near a
sphere}\label{sec:particlesphere}

\begin{figure}[tb]
  \includegraphics[width=1.5in]{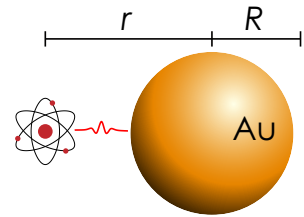}
  \caption{The geometry considered: a quantum particle prepared in
  eigenstate $|n\rangle$ outside a gold sphere.}
  \label{fig:geometry}
\end{figure}

As depicted in Fig.~\ref{fig:geometry}, we consider a particle at
distance $r$ from the center of a sphere of radius $R$ and
permittivity $\varepsilon=\varepsilon(\omega)$.
%%%%%%%%%%%%%%%  F I G U R E %%%%%%%%%%%%%%%%%%%%%%%%%%%%%%%%%%%%%%%%%
%%%%%%%%%%%%%%%%%%%%%%%%%%%%%%%%%%%%%%%%%%%%%%%%%%%%%%%%%%%%%%%%%%%%%%
The dyadic Green's function leads to \cite{buhmann04}
\begin{align}
&\dGamma_\omega(\vc{r})
= \frac{ix}{4\pi r^3}\sum_{l=1}^\infty (2l+1)\Bigl\{ x^2 \rs (\phi x)
 \hl(x)^2 \notag \\
&+\rp(\phi x) \Bigl[ l(l+1) \hl(x)^2  + \thl(x)^2 \Bigr]
 \Bigr\}\label{trGamma}
\end{align}
where $\hl(x)$ is the spherical Hankel function of the first kind,
$\thl(x)$ [and for future reference, $\tjl(x)$] is shorthand for
\be\label{shorthand}
\thl(x) = [x\hl(x)]'; ~~  
  \tjl(x) = [x j_l(x)]'
\ee
[$j_l(x)$: spherical Bessel function of the first kind].
For convenience we are using the dimensionless distance and size
parameters $x=r\omega /c$ and $0<\phi=R/r<1$, recall
Eqs.~\eqref{xdef} and \eqref{phidef}. The reflection coefficients for
TE and TM-polarized waves read
\begin{subequations}\label{reflcoeffs}
 \begin{align}
 \rs(z) =&
  -\frac{\tjl(z)j_l(\sqe z) - \tjl(\sqe z)j_l(z)}{\thl(z)j_l(\sqe z)
  - \tjl(\sqe z)\hl(z)}; \label{rs}\\
 \rp(z) =& -\frac{\varepsilon \tjl(z)j_l(\sqe z)
  - \tjl(\sqe z)j_l(z)}{\varepsilon\thl(z)j_l(\sqe z)
  - \tjl(\sqe z)\hl(z)}.\label{rp}
  \end{align}
\end{subequations}

In the following we assume both the particle--center separation
and the sphere size to be non-retarded, $\phi x\le x\ll 1$. In
addition, we perform the perfect-conductor limit $|\varepsilon|\gg 1$.
To leading order, the two limits commute: Taking the perfect-conductor
limit first, the reflection coefficients reduce to
\begin{gather}\label{rTEPC}
  \rs(\phi x) \buildrel{\varepsilon\to \infty}\over{\longrightarrow}
  \rsPC(\phi x) = -\frac{j_l(\phi x)}{\hl(\phi x)};\\
\label{rTMPC}
  \rp(\phi x) \buildrel{\varepsilon\to \infty}\over{\longrightarrow}
  \rpPC(\phi x) = -\frac{\tjl(\phi x)}{\thl(\phi x)},
\end{gather}
see the asymptotes~\eqref{jtoinfty} and \eqref{htoinfty} in
App.~\ref{app}. Using the expansions~\eqref{jto0} and \eqref{hto0},
they further simplify to
\begin{gather}\label{rTE0PC}
  \rsPC(\phi x) \buildrel{\phi x\to 0}\over{\longrightarrow}
  r^{\mathrm{TE},\PC}_{l,0}(\phi x) =
 -\frac{\rmi(\phi x)^{2l+1}}{(2l\!+\!1)!!(2l\!-\!1)!!}\,;\\
\label{rTM0PC}
 \rpPC(\phi x) \buildrel{\phi x\to 0}\over{\longrightarrow}
  r^{\mathrm{TM},\PC}_{l,0}(\phi x)
  = \frac{l\!+\!1}{l}\,\frac{\rmi(\phi x)^{2l+1}}
 {(2l\!+\!1)!!(2l\!-\!1)!!}
\end{gather}
in the non-retarded limit. Here, $(2l+1)!!=1\cdot 3\cdots (2l+1)$.

In contrast, when taking the non-retarded limit first, one finds
\begin{gather}\label{rTE0}
  \rs(\phi x) \buildrel{\phi x\to 0}\over{\longrightarrow}
  r^{\mathrm{TE}}_{l,0}(\phi x) =
 (\varepsilon\!-\!1)\,
  \frac{\rmi(\phi x)^{2l+3}}{(2l\!+\!3)!!(2l\!+\!1)!!}\,;\\
\label{rTM0} \rp(\phi x) \buildrel{\phi x\to 0}\over{\longrightarrow}
  r^{\mathrm{TM}}_{l,0}(\phi x) =
  \frac{(l\!+\!1)(\varepsilon\!-\!1)\rmi(\phi x)^{2l+1}}
 {(l\varepsilon\!+\!l\!+\!1)
 (2l\!+\!1)!!(2l\!-\!1)!!}\,,
\end{gather}
and subsequently
\begin{gather}
r^{\mathrm{TE}}_{l,0}(\phi x)
 \buildrel{\varepsilon\to\infty}\over{\longrightarrow}
  \varepsilon\,
  \frac{\rmi(\phi x)^{2l+3}}{(2l+3)!!(2l+1)!!}\,;\\
r^{\mathrm{TM}}_{l,0}(\phi x)
 \buildrel{\varepsilon\to\infty}\over{\longrightarrow}
 \frac{l+1}{l}\,
  \frac{\rmi(\phi x)^{2l+1}}{(2l+1)!!(2l-1)!!}\,.
\end{gather}
While the $\mathrm{TM}$-coefficient takes the same form regardless of
the order of the limits, the $\mathrm{TE}$-coefficient yields
different results, depending on which of the two limits is performed
first.

However, the Green tensor in the non-retarded limit is dominated by
$\rp$. Substituting the results for the reflection coefficients into
\eqref{trGamma}, using the approximation~\eqref{hto0} from
App.~\ref{app} and retaining only the leading order in $x$, one finds
\begin{align}
\dGamma_0^\PC(\vc{r})
=&\frac{1}{4\pi r^3}\sum_{l=1}^\infty (2l+1)(l+1)\phi^{2l+1}\notag\\
=&\frac1{4\pi r^3}\frac{\phi^3 (6-3\phi^2 +
\phi^4)}{(1-\phi^2)^3}\,.\label{trGammaPC0}
\end{align}
With this result, the temperature-invariant CP potential~\eqref{Tinv}
reads
\be
U_0(\vc{r})
=-\frac{|\vc{d}|^2}{24\pi\varepsilon_0 r^3}\,f(\phi)\label{UspherePC0}
\ee
with 
\be
f(\phi)
=\frac{\phi^3 (6-3\phi^2 + \phi^4)}{(1-\phi^2)^3}
\to\begin{cases}6\phi^3&\mbox{for }\phi\to 0,\\
\displaystyle
\frac{1}{2(1-\phi)^3}
&\mbox{for }\phi\to 1.\end{cases}
\label{scaling}
\ee
This is in agreement with the zero-temperature potential as found in
Ref.~\cite{taddei10} for a perfectly conducting sphere in the
non-retarded limit on the basis of image-charge techniques. As
discussed in Ref.~\cite{buhmann10}, the atom-sphere geometry is a
particular example of a two-parameter geometry, conveniently
described by a scaling function $f(\phi)$.

\begin{figure}[tb]
  \includegraphics[width=3.3in]{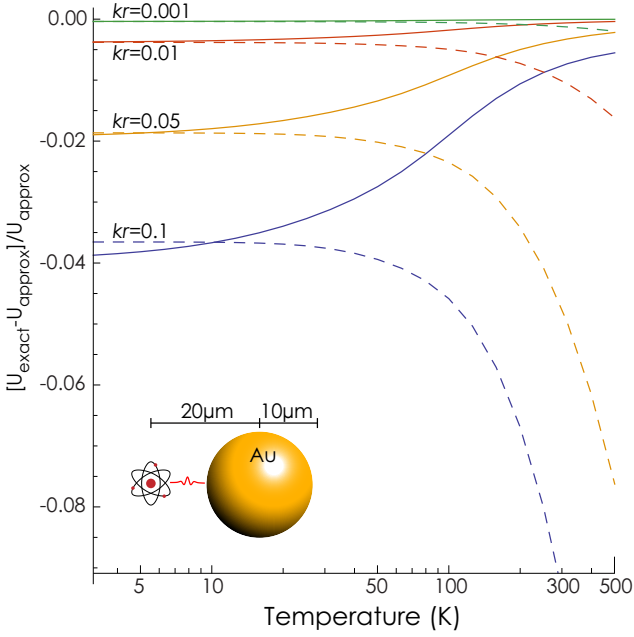}
  \caption{Numerical comparison of the exact CP potential $U(T)$ of 
the contribution from a transition of energy $\hbar c k$ for a
particle outside a gold sphere with
$U_\text{approx}$ being respectively 
the $T$-independent result $U_0$ for a perfect
conductor in the non-retarded limit 
(dashed line) and 
the approximate
expression Eq.~\eqref{finalresult} including the linear temperature 
correction (solid line). 
Parameter values: $R = \half r =10\mu$m.
}
  \label{fig:spheregraph}
\end{figure}

\begin{figure}[tb]
\includegraphics[width=3.3in]{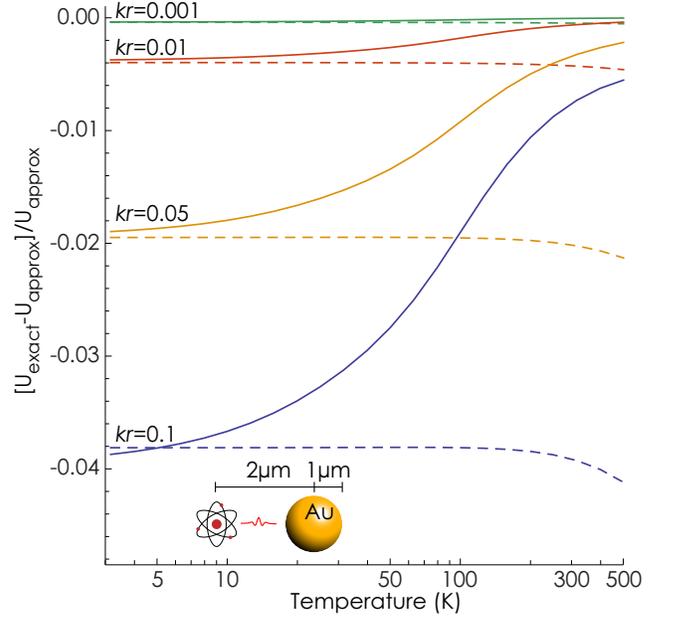}
\caption{
Same as figure \ref{fig:spheregraph}, but with a smaller
gold sphere:
$R = \half r = 1\mu$m.}
  \label{fig:smallspheregraph}
\end{figure}

The accuracy of the simplest, temperature-independent approximation,
Eq.~\eqref{UspherePC0}, is demonstrated numerically in
Fig.~\ref{fig:spheregraph} (dashed lines) for a ground-state two-level
particle outside a gold sphere. The permittivity of the sphere has
been described by a Drude model
$\varepsilon(\omega)=1-\omega_P^2/[\omega(\omega+i\gamma)]$ with
parameters $\omega_P=9$eV and $\gamma=35$meV. For comparison, the same
situation but with a smaller sphere is shown in
Fig.~\ref{fig:smallspheregraph}. We see that Eq.~\eqref{UspherePC0}
yields a very good approximation. The exact potential is slightly
smaller by at most $5\%$ for $r\omega/c=0.1$. Recall that the leading
correction in the high-temperature limit has opposite signs for
ground-state and excited atoms. For an excited two-level atom, we
would hence find that the exact potential is slightly larger than its
approximation~\eqref{UspherePC0}. In the following, we derive
analytical expressions for the leading temperature-dependent
corrections to Eq.~\eqref{UspherePC0}
providing an even much improved approximation at higher temperatures.

%%%%%%%%%%%%%  S U B S E C T I O N %%%%%%%%%%%%%%%
\subsection{Correction from retardation}\label{sec_ret}

When considering the full thermal CP potential~(\ref{eq:nonres}) and
(\ref{eq:res}) using the Green's function of Eq.~\eqref{trGamma}, the
non-retarded and perfect-conductor limits do not commute. We have
$\phi x\le x\ll 1$ and $|\varepsilon|\gg 1$, which is compatible with
both large and small values of $|\sqrt{\varepsilon}|\phi x$. For a
metal sphere at typical experimental distances of order micrometers
and $x\sim 0.01-0.001$, we have $\im\sqrt{\varepsilon}\phi x\gg 1$,
meaning that the perfect-conductor limit has to be applied first. The
opposite limit $|\sqrt{\varepsilon}|\phi x\ll 1$ may be realised for
dielectrics whose permittivity tends to some moderate electrostatic
value, in which case the nonretarded limit would have to be performed
first. We briefly consider this case in Appendix \ref{app:dielectric}.

The correction from retardation effects is found by using the perfect
conductor values~\eqref{rTEPC} and \eqref{rTMPC} of the reflection
coefficients and expanding $\dGamma_\omega^\PC(\vc{r})$ as given by
Eq.~\eqref{trGamma} in powers of $x$, the leading correction term
being of order $x^2$. We obtain such quadratic corrections from three
sources: (A) from the $\mathrm{TM}$-mode reflection coefficient
$\rpPC$; (B) from $\mathrm{TM}$-mode propagators $\hl(x)^2$ and
$\thl(x)^2$; and (C) from the leading-order $\mathrm{TE}$-mode
contribution. The technical details of the small $x$ expansions of the
different cases are found in App.\ref{appA}.
%%%%%%%%%%%%%%%  F I G U R E %%%%%%%%%%%%%%%%%%%%%%%%%%%%%%%%%%%%%%%%%
%%%%%%%%%%%%%%%%%%%%%%%%%%%%%%%%%%%%%%%%%%%%%%%%%%%%%%%%%%%%%%%%%%%%%%

%%%%%%%%%%%%%%%  F I G U R E %%%%%%%%%%%%%%%%%%%%%%%%%%%%%%%%%%%%%%%%%
%%%%%%%%%%%%%%%%%%%%%%%%%%%%%%%%%%%%%%%%%%%%%%%%%%%%%%%%%%%%%%%%%%%%%%

As shown therein, correction (A) takes the form
\begin{align}
  \rpPC(\phi x) =& r^{\mathrm{TM},\PC}_{l,0}(\phi x) \left\{1
  - \frac{(\phi
 x)^2}{2}\left[\frac{l+3}{(2l+3)(l+1)}  \right.\right. \notag \\
  &\left.\left.+\frac{l-2}{l(2l-1)}\right]+... \right\}\label{refret}
\end{align}
with $r^{\mathrm{TM},\PC}_{l,0}$ being given by Eq.~\eqref{rTM0PC}.
The correction (B) is found to be
\begin{align}
  r^{\mathrm{TM},\PC}_{l,0}(\phi x)&\Bigl[ l(l+1) \hl(x)^2 + \thl(x)^2
\Bigr]\notag \\
  &= \frac{(l+1)\phi^{2l+1}}{ix}\left[1
+ \frac{x^2}{2l+1} + ... \right].\label{propagcorr}
\end{align}
Finally, the correction (C) from the $\mathrm{TE}$-mode takes the
simple form
\be\label{s-corr}
\rsPC(\phi x) \hl(x)^2 = -\frac{\phi^{2l+1}}{ix(2l+1)} + ...
\ee

Substituting these corrections into Eq.~\eqref{trGamma}, one finds
\begin{align}
\Delta\dGamma^\text{ret.}_\omega =
\frac{x^2}{4\pi r^3}\sum_{l=1}^\infty
\biggl\{&l\phi^{2l+1}-(2l+1)\biggl[\frac{l+3}{2l+3}\notag\\
 &+\frac{(l+1)(l-2)}{(2l-1)l}\biggr]\frac{\phi^{2l+3}}{2}\biggr\}
\end{align}
The sum may be carried out in closed form by splitting the expressions
into partial fractions. One finds
\begin{align}
\Delta\dGamma^\text{ret.}_\omega =& \frac{x^2}{8\pi r^3}\bigl[
3(1+3\phi^4)\mathrm{artanh}\phi-\phi(3-\phi^2)\notag \\
  &+2\phi^3 \log(1-\phi^2)\bigr].\label{retcorr}
\end{align}

Substituting this result into Eq.~\eqref{DUi}, we obtain the
retardation correction
\be
\Delta U_\text{ret.}(\vc{r})
=\frac{|\vc{d}|^2x^2}{24\pi\varepsilon_0 r^3}
 \left(\frac{\kB T}{\hbar\omega}-\frac{1}{2}\right)
 \,g_\text{ret.}(\phi)\label{DUSphereret}
\ee
with scaling function
\begin{align}
&g_\text{ret.}(\phi)\notag\\
&=3(1+3\phi^4)\mathrm{artanh}\phi-\phi(3-\phi^2)
  +2\phi^3 \log(1-\phi^2)\notag\\
&\to\begin{cases}2\phi^3&\mbox{for }\phi\to 0,\\
\displaystyle-4\log(1-\phi)&\mbox{for }\phi\to 1.\end{cases}
\label{scalingret}
\end{align}
Its relative contribution~\eqref{DUret} is given by
\begin{equation}\label{DUSphererelret}
\frac{\Delta U_\mathrm{ret.}(\vc{r})}{U_0(\vc{r})}
  =-\left(\frac{\kB T}{\hbar\omega}-\frac{1}{2}\right)x^2\,
 \frac{g_\text{ret.}(\phi)}{f(\phi)}\,.
\end{equation}

%%%%%%%%%%%%%  S U B S E C T I O N %%%%%%%%%%%%%%%
\subsection{Correction from imperfect reflection}

The leading order corrections to the ideal reflection coefficients are
calculated in App.~\ref{appB} and have the form
\begin{subequations}\label{rcorr}
\begin{align}
\rs(\phi x)=&r^{\mathrm{TE},\PC}_{l,0}(\phi x)
\left[1 - \frac{i(2l+1)} {\sqe \phi x}+...\right];\\
  \rp(\phi x)=&r^{\mathrm{TM},\PC}_{l,0}(\phi x)
\left[1 + \frac{i\phi x}{\sqe}\,
\frac{2l+1}{l(l+1)}+...\right]
\end{align}
\end{subequations}
for $x\ll 1$, with $r^{\mathrm{TE},\PC}_{l,0}$ and
$r^{\mathrm{TM},\PC}_{l,0}$ given by Eqs.~\eqref{rTE0PC} and
\eqref{rTM0PC}. Substituting these results into Eq.~\eqref{trGamma}
and using the leading-order small argument expansions in
Eqs.~\eqref{propagcorr} and \eqref{s-corr}, we find
\begin{align}
  \Delta&\dGamma_\omega^\text{refl.}=\frac{ix}{4\pi r^3\sqe}
 \sum_{l=1}^\infty(2l+1)
\left[1+\frac{2l+1}{l}\phi^2\right]
\phi^{2l}\notag \\
  =& \frac{ix\phi^2}{4\pi r^3\sqe}\left[\frac{3+7\phi^2 - 4\phi^4}
{(1-\phi^2)^2}-\log(1-\phi^2) \right].\label{reflcorr}
\end{align}

Inserting this into Eq.~\eqref{DUi}, the reflectivity correction is
found to be
\be
\Delta U_\text{refl.}(\vc{r})
=\frac{|\vc{d}|^2x}{24\pi\varepsilon_0 r^3}
\re\biggl(\frac{\rmi}{\sqe}\biggr)
 \left(\frac{\kB T}{\hbar\omega}-\frac{1}{2}\right)
 \,g_\text{refl.}(\phi)\label{DUSphererefl}
\ee
with scaling function
\begin{align}
&g_\text{refl.}(\phi)
=2\phi^2\left[\frac{3+7\phi^2 - 4\phi^4}
{(1-\phi^2)^2}-\log(1-\phi^2) \right]\notag\\
&\to\begin{cases}6\phi^2&\mbox{for }\phi\to 0,\\
\displaystyle\frac{3}{(1-\phi)^2}&\mbox{for }\phi\to 1.\end{cases}
\label{scalingrfl}
\end{align}
Its relative contribution~\eqref{DUret} reads
\begin{equation}\label{DUSphererelrefl}
\frac{\Delta U_\mathrm{refl.}(\vc{r})}{U_0(\vc{r})}
  =-\left(\frac{\kB T}{\hbar\omega}-\frac{1}{2}\right)x
 \re\biggl(\frac{\rmi}{\sqe}\biggr)\,
 \frac{g_\text{refl.}(\phi)}{f(\phi)}\,.
\end{equation}

%%%%%%%%%%%%%%%%%%%%%%%%%%%%%%%%%%%%%%%%%%%%%%%%%%
%%%%%%%%%%%%%  S U B S E C T I O N %%%%%%%%%%%%%%%
%%%%%%%%%%%%%%%%%%%%%%%%%%%%%%%%%%%%%%%%%%%%%%%%%%
\subsection{Discussion and comparison}\label{sec:discussion}

Combining the $T$-invariant result~\eqref{Tinv} with the retardation
and reflectivity corrections~\eqref{DUSphereret} and
\eqref{DUSphererefl}, we obtain the weakly temperature-dependent CP
potential \eqref{finalresult}, as stated in the introduction.
The quality of this analytic high-temperature result is demonstrated
in Figs.~\ref{fig:spheregraph} and \ref{fig:smallspheregraph} (solid
lines), where we compare it with the result of an exact numerical
calculation for
the contribution $U_{nk}$ [see Eq.~\eqref{Un}] from an upward internal 
energy transition of a particle
outside a
gold sphere. One sees that the analytic result is an excellent
approximation for temperatures $T>200\,\mathrm{K}$. 
Notice in particular that for $x$ of the order $0.1$ and lower, 
Eq.~\eqref{finalresult} is an excellent approximation (better than $1\%$) 
at $T\approx 300$K, where most experiments are performed.

While the T-invariant first term in Eq.~\eqref{finalresult} is quite 
adequate for very small retardation values (such as $x=0.001$, 
typical of Rydberg atoms), the full expression is much better as $x$
increases to about $0.1$. This could be the case for certain cold
polar molecules. For example, with LiH molecules \cite{ellingsen09},
$x=0.1$ corresponds to $r=15\mu$m for the dominant, rotational
transition which is not an atypical situation. In this case 
the $T$-independent term is off by almost $10\%$ at $300$K with the
$10$ micron sphere, but is still better than $1\%$ when the
correction is included.
In general, the T-invariant first term in Eq.~\eqref{finalresult}
becomes a worse approximation at higher temperatures, with the error
increasing without bounds as the environment temperature rises. In
contradistinction, our approximation~\eqref{finalresult} becomes
better and better at high temperatures and its error remains bounded
throughout.

\begin{figure}[tb]
\includegraphics[width=3in]{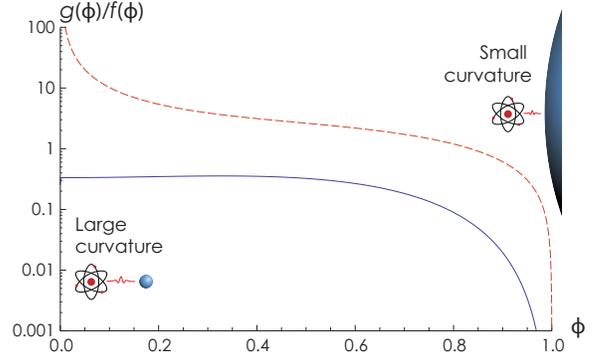}
\caption{
Ratios of the scaling functions $g_\mathrm{ret.}(\phi)/f(\phi)$
(solid line) and $g_\mathrm{refl.}(\phi)/f(\phi)$ (dashed line) that
govern the impact of retardation and finite reflectivity on the
thermal CP potential.}
  \label{fig:scaling}
\end{figure}

According to Eqs.~\eqref{DUSphererelret} and \eqref{DUSphererelrefl},
the relative contributions from retardation and finite reflectivity
are governed by the ratios $g_i(\phi)/f(\phi)$ of the scaling
functions as given by Eqs.~\eqref{scaling},\eqref{scalingret} and
\eqref{scalingrfl}. These ratios are depicted in
Fig.~\ref{fig:scaling}.
%%%%%%%%%%%%%%%  F I G U R E %%%%%%%%%%%%%%%%%%%%%%%%%%%%%%%%%%%%%%%%%
%%%%%%%%%%%%%%%%%%%%%%%%%%%%%%%%%%%%%%%%%%%%%%%%%%%%%%%%%%%%%%%%%%%%%%
The figure shows that both contributions strongly depend on the
curvature of the sphere as parametrised by $\phi$. The retardation
contribution takes a value $1/3$ for a strongly curved sphere and
stays approximately constant for $\phi\lesssim 0.5$. In the limit of a
flat surface, it rapidly falls off as $-4(1-\phi)^3\log(1-\phi)$. The
reflectivity correction grows as $1/\phi$ in the limit of a strongly
curved sphere and falls off gently as $6(1-\phi)$ in the limit of a
flat surface.

The ratio $g_\mathrm{refl.}(\phi)/f(\phi)$ is greater than
$g_\mathrm{ret.}(\phi)/f(\phi)$ by at least an order of magnitude for
all curvatures. One has to bear in mind, however, that the
reflectivity correction carries an additional factor
$\re(\rmi/\sqe)/x\ll 1$. For a metal described by a Drude model with 
$\omega,\gamma\ll\omega_P$, one finds
\begin{align}
\re\bigl[\rmi/\sqrt{\varepsilon(\omega)}\bigr]
&=\omega_P^{-1}
 \left[\half(\sqrt{\omega^2+\gamma^2}+|\omega|)
 |\omega|\right]^\frac1{2}\notag\\
&\simeq\begin{cases}\sqrt{\omega\gamma/2}/\omega_P
 &\mbox{for }\omega\ll\gamma,\\
 \omega/\omega_P&\mbox{for }\omega\gg\gamma.\end{cases}
\end{align}
Depending on the actual value of $\re(\rmi/\sqe)/x\ll 1$ for a given
molecule and material, either one of the reflectivity or retardation
may dominate for given curvatures. However, the asymptotic behaviour
observed in Fig.~\ref{fig:scaling} shows that the reflectivity
correction will always dominate in the limits of small or large
curvature.

%%%%%%%%%%%%%%%%%%%%%%%%%%%%%%%%%%%%%%%%%%%%%%%%%%
%%%%%%%%%%%%%  S U B S E C T I O N %%%%%%%%%%%%%%%
%%%%%%%%%%%%%%%%%%%%%%%%%%%%%%%%%%%%%%%%%%%%%%%%%%
\section{Summary}

We have studied the
temperature-dependent
CP potential of a particle 
strongly out of thermal equilibrium
near a metal sphere,
and have derived a simple approximate expression in closed form for
the interaction potential. The approximation is valid at the $1\%$ or
better for Rydberg atoms and cold polar molecules at all temperatures
for typical experimental length scales.
We have assumed both the particle--sphere distance and the particle's
transition frequency to be small enough so that retardation and
imperfect reflectivity present only small perturbations to the
temperature-independent result. Such is typically the case in
experimental set-ups in which cold polar molecules or Rydberg atoms
are used, whose interaction potential is dominated by long wavelength
transitions for which the non-retarded regime extends to tens and
hundreds of micrometers, respectively.

In recent publications it has been shown that the Casimir--Polder
potential acting on a particle prepared in an eigenstate at a
non-retarded distance from a macroscopic body can be virtually
independent of the surrounding temperature. This is the case for the
geometry considered, and the error made in approximating the
interaction as independent of temperature from zero to room
temperature is only a few percent for sufficiently non-retarded 
interaction. The small temperature-dependent corrections to the
potential of a metal body have been identified to stem from
retardation and imperfect reflectivity and we have analysed these
separately and discussed the relative importance of each with respect
to the other. Our results show that reflectivity is the dominant
correction for very large or small curvatures, while intermediate
curvatures may be governed by either retardation and reflectivity
corrections, depending on particle and material.

The perturbative method employed in this investigation is equally
well suited for the study of more complicated geometries. Again, it
promises physical insights that are hard or even impossible to gain by
numerical means.

We thank Ho Trung Dung for discussions. This work was supported by the
UK Engineering and Physical Sciences Research Council. Support from
the European Science Foundation (ESF) within the activity `New Trends
and Applications of the Casimir Effect' is gratefully acknowledged.

%%%%%%%%%%%%%%%%%%%%%%%%%%%%%%%%%%%%%%%%%%%%%%%%
%%%%%%%%%%%%%%% A P P E N D I X %%%%%%%%%%%%%%%%
%%%%%%%%%%%%%%%%%%%%%%%%%%%%%%%%%%%%%%%%%%%%%%%%
\appendix

\section{Limits and their leading corrections}\label{app}

To calculate the CP potential in the perfect-conductor and
non-retarded limits, we need to approximate the spherical Bessel
functions for small and large arguments. Using the asymptotes from
\S10 of Ref.~\cite{abramowitz}
\begin{gather}\label{jtoinfty}
 j_l(x) \approx \frac{\sin(x-l\pi/2)}{x}
 \quad\mbox{for }x\gg 1;\\
\label{htoinfty}
\hl(x) \approx \frac{(-\rmi)^{l+1}\rme^{\rmi(x-n\pi/2)}}{x}
 \quad\mbox{for }x\gg 1,
\end{gather}
we easily find the perfect-conductor limits~\eqref{rTEPC} and
\eqref{rTMPC}. 

The non-retarded limits~\eqref{rTE0PC}, \eqref{rTM0PC}, \eqref{rTE0}
and \eqref{rTM0} can be found by using the expansions
\cite{abramowitz}
\begin{gather}\label{jto0}
 j_l(x) = \frac{x^l}{(2l+1)!!}
\left[1-\frac{x^2}{2(2l+3)}
+...\right]\\
\label{hto0}
\hl(x) = -\frac{\rmi (2l-1)!!}{x^{l+1}}
\left[1+\frac{x^2}{2(2l-1)}
+...\quad\right]
\end{gather}
for $x\ll 1$ with $(2l+1)!!=1\cdot 3 \cdot 5 \cdots (2l+1)$. Note that
the next-to-leading order expansion is only needed for \eqref{rTE0},
where the leading-order term vanishes.

%%%%%%%%%%%%%%%%%%%%%%%%%%%%%%%%%%%%%%%%%%%%%%%%
\subsection{Retardation corrections}\label{appA}

To calculate the retardation correction to the perfect-conductor
$TM$-mode reflection coefficients~\eqref{rTMPC}, we use the expansions
\eqref{jto0} and \eqref{hto0}, which together with the
definitions~\eqref{shorthand} lead to
\begin{subequations}
\begin{align}
  \tjl(x) = & \frac{(l+1)x^l}{(2l+1)!!}\left[1 -
\frac{x^2}{2}\frac{l+3}{(2l+3)(l+1)}+...\right]\\
  \thl(x)=& \frac{il(2l-1)!!}{x^{l+1}}\left[1 +
\frac{x^2}{2}\frac{l-2}{l(2l-1)}+...\right]\label{dhto0}
\end{align}
\end{subequations}
for $x\ll 1$. This immediately yields Eq.~(\ref{refret}).

The retardation correction from the propagator factors are found from
expansions
\begin{subequations}
\begin{align}
 \hl(x)^2 =& -\frac{[(2l-1)!!]^2}{x^{2l+2}}\left[1 +
\frac{x^2}{2l-1}+...\right];\label{hexp}\\
  \thl(x)^2 =& -\frac{l^2[(2l-1)!!]^2}{x^{2l+2}}\left[1 + \frac{x^2
(l-2)}{l(2l-1)}+...\right]
\end{align}
\end{subequations}
which follow from Eqs.~\eqref{hto0} and \eqref{dhto0}.
Combining these results with $r^{\mathrm{TM},\PC}_{l,0}$ as given by
Eq.~\eqref{rTM0PC}, we arrive at Eq.~\eqref{propagcorr}.

Expansion~\eqref{hexp} moreover combines with
$r^{\mathrm{TE},\PC}_{l,0}$ as given by Eq.~\eqref{rTE0PC} to result
in Eq.~\eqref{s-corr}.

%%%%%%%%%%%%%%%%%%%%%%%%%%%%%%%%%%%%%%%%%%%%%%%%
\subsection{Finite reflectivity correction}\label{appB}

In order to expand the reflection coefficients~\eqref{reflcoeffs} in
powers of $\varepsilon^{-1}$, we rewrite them as
\begin{align*}
\rp(\phi x)
=& \rpPC(\phi x)\frac{1-AJ/\varepsilon}{1-AH/\varepsilon}\,;\\
\rs(\phi x) =& \rsPC(\phi x) \frac{1-1/(AJ)}{1-1/(AH)}
\end{align*}
where
\[
  A = \frac{\tjl(\sqe\phi x)}{j_l(\sqe\phi x)}\,; ~~ 
  J = \frac{j_l(\phi x)}{\tjl(\phi x)}\,; ~~
  H=\frac{\hl(\phi x)}{\thl(\phi x)}\,.
\]
With the assumption $\im\{\sqe\}\phi x\gg 1$,
the asymptote~\eqref{jtoinfty} leads to
\begin{align*}
\frac{\tjl(\sqe\phi x)}{j_l(\sqe\phi x)}\approx& \sqe \phi x\cot(\sqe
\phi x-{\textstyle \frac{l\pi}{2}})\approx -i\sqe \phi x. % \\
\end{align*}
Using the asymptotes~\eqref{jto0} and \eqref{hto0}, we further have
\[
 \frac{j_l(\phi x)}{\tjl(\phi x)}\approx \frac1{l+1}; ~~~
\frac{\hl(\phi x)}{\thl(\phi x)}\approx -\frac1{l}
\]
for $x\ll 1$. Combining these results and retaining only the
next-to-leading order in $x$, one easily obtains Eqs.~\eqref{rcorr}.

\section{Casimir-Polder expression for a dielectric sphere}
\label{app:dielectric}

An expression for the CP potential on a particle near a dielectric
sphere can be readily derived using the same methods as elsewhere in
this article. Although it may be written in closed form in terms of
hypergeometric functions, this hardly constitutes a simplification,
and we will only give the expression as an infinite sum. The sphere's
dielectric constant $\varepsilon$ is now no longer assumed large, so
that $|\sqrt{\varepsilon}|x\ll 1$ is assumed. The TE contribution is
now of order $x^4$ and can be neglected. Since we are considering
frequencies $\omega_{kn}$ which are typically small on an optical
scale, we assume $\ep(\omega_{kn})=\ep(0)=\ep$. The result is
 \begin{align}\label{Udiel}
 U_{nk}^\text{diel.}\approx&-\frac{|\vc{d}_{kn}|^2(\ep-1)}{
 24\pi\varepsilon_0 r^3}\sum_{l=1}^\infty
  \frac{l(l+1)(2l+1)\phi^{2l+1}}{\varepsilon l+l+1}\notag \\
  &\times\Bigl\{1+2x^2\Bigl(\frac{\kB T}{\hbar
 \omega_{kn}}+\frac1{2}\Bigr)\Bigl[\frac{1}{2l+1}\notag \\
  &-\frac{\phi^2 (2l+1)}{(2l-1)(2l+3)}\frac{\ep(l-2)+l+1}{\ep
 l+l+1}\Bigr]\Bigr\}.
 \end{align}
Eq~\eqref{Udiel} is the counterpart of \eqref{finalresult} for the
case of a dielectric sphere. 

%%%%%%%%%%%%%%%%%%%%%%%%%%%%%%%%%%%%%%%%%%
%%%%%%%%%%%%%%%%  F I G U R E
%%%%%%%%%%%%%%%%%%%%%%%%%%%%%%%%%%%%%%%%%%
\begin{figure}[!t!]
  \includegraphics[width=3.3in]{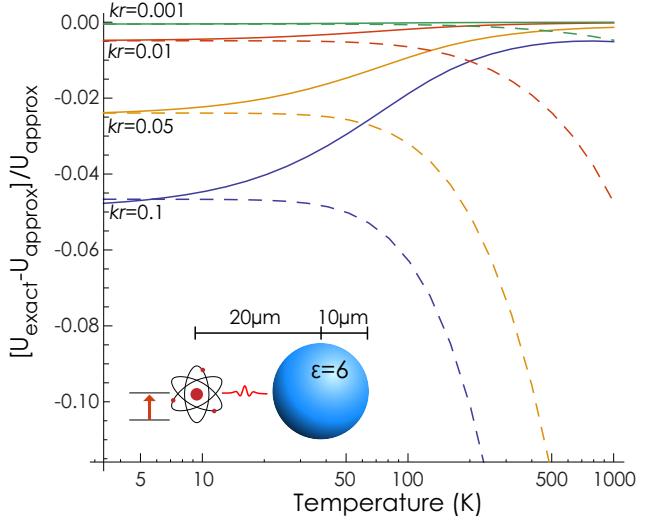}
  \caption{Numerical comparison of the exact CP potential $U(T)$ of a  
  two-level model particle (transition energy $\hbar c k$) outside a 
  dielectric sphere with the approximation Eq.~\eqref{Udiel} with
  (solid line) and without (dashed lines) the correction term in the
  curly brackets. Parameter values: $R =\half r = 10\mu$m and
  $\ep=6$.}
\label{fig:dielectric}
\end{figure}

As for the case of a metal sphere, the approximation is good to at
least the $1\%$ level at all temperatures when $x\lesssim 0.01$, and
becomes significantly better for $T\gg \hbar \omega/\kB$ (about $11$K
for $x=0.01$ with the numbers in figure \ref{fig:dielectric}).

When assuming as we have that $\varepsilon$ does not vary appreciably
between frequencies $0$ and $\omega_{kn}$, the leading order
correction is of order $x^2$. It is noteworthy that the T-independent
term alone (without the correction in the curly braces of
Eq.~\eqref{Udiel}) is about as good an approximation as the
corresponding expression is for a metal sphere. This conclusion would
not hold, however, if the dielectric material has resonances at
frequencies in the order of $\omega$, which might lie in the
microwave or far infrared regime, in which case $\Delta
\Gamma_\omega$ may no longer be small compared to $\Gamma_0$ (see
Section \ref{sec:generalformalism}).

%%%%%%%%%%%%%%%%%%%%%%%%%%%%%%%%%%%%%%%%%%%%%%%%%%%%%%%%%%%%%%%%%%%


\begin{thebibliography}{99} 

\bibitem{casimir48}
H.\ B.\ G.\ Casimir and D.~Polder, Phys.\ Rev.\ \textbf{73}, 360
(1948).
\bibitem{gallagher94}
T.~F.\ Gallagher, \textit{Rydberg Atoms} (Cambridge University Press,
1994).
\bibitem{kubler10} H.~K\"ubler, J.~P.\ Shaffer, T.~Baluktsian,
R.~L\"ow, and T.~Pfau, Nature Photonics {\bf 4}, 112 (2010).
\bibitem{tauschinsky10} A.~Tauschinsky, R.~M.~T.\ Thijssen,
S.~Whitlock, H.~B.\ van Linden van den Heuvell, and R.~J.~C.\ Spreeuw,
Phys.\ Rev.\ A {\bf 81}, 063411 (2010).
\bibitem{sorensen04} A.~S.\ S\o rensen, C.~H.\ van der Wal, L.~I.\
Childress, and M.~D.\ Lukin, Phys.\ Rev.\ Lett {\bf 92}, 063601
(2004).
\bibitem{hyafil04} P.~Hyafil, J.~Mozley, A.~Perrin, J.~Tailleur,
G.~Nogues, M.~Brune, J.~M.\ Raimond, and S.~Haroche, Phys.\ Rev.\ Lett
{\bf 93}, 103001 (2004).
\bibitem{petrosyan09} D.~Petrosyan, G.~Bensky, G.~Kurizki, I.~Mazets,
J.~Majer, and J.~Schmiedmayer, Phys.\ Rev.\ A {\bf 79}, 040304(R)
(2009).
\bibitem{vandemeerakker08}
S.~Y.~T.~van~de~Meerakker, H.~L.~Bethlem, and G.~Meijer, Nature
Physics \textbf{4}, 595 (2008).
\bibitem{carr09} L.~D.\ Carr, D.~DeMille, R.~V.\ Krems, and J.~Ye, New
J.\ Phys.\ {\bf 11}, 055049 (2009).
\bibitem{bell09} M.~T.\ Bell and T.~P.\ Softley, Mol.\ Physics {\bf
107}, 99 (2009).
\bibitem{meek09} S.~A.\ Meek, H.~Conrad, and G.~Meijer, Science {\bf
324}, 1699 (2009).
\bibitem{ashkin70} 
A.~Ashkin, Phys.\ Rev.\ Lett.\ {\bf 24}, 156 (1970).
\bibitem{chen11} 
J.~Chen, J.~Ng, Z.~Lin, and C.~T.\ Chan, Nature Photonics {\bf 5}, 531
(2011).
\bibitem{kawata92} S.~Kawata and T.~Sugiura, Opt.\ Lett {\bf 17}, 772
(1992).
\bibitem{min03} S.~M.\ Spillane, T.~J.\ Kippenberg, and K.~J.\ Vahala,
Nature {\bf 415}, 621 (2002).
\bibitem{Haakh}
H.~Haakh, F.~Intravaia, C.~Henkel, S.~Spagnolo, R.~Passante, B.~Power,
and F.~Sols, Phys. Rev. A \textbf{80}, 062905 (2009).
\bibitem{buhmann08b} S.Y.~Buhmann, M.R.~Tarbutt, S.~Scheel, and
E.A.~Hinds, Phys. Rev. A \textbf{78}, 052901 (2008).
\bibitem{crosse10}
J.\ A.\ Crosse,  S.\ {\AA}.\ Ellingsen, K.~Clements, S.\ Y.\ Buhmann,
and S.\ Scheel, Phys. Rev. A {\bf 82}, 010901(R) (2010); Erratum
{\it ibid.}\ {\bf 82}, 029902(E) (2010).
\bibitem{buhmann08}
S.~Y.\ Buhmann and S.~Scheel, Phys.\ Rev.\ Lett.\ \textbf{100}, 253201
(2008).
\bibitem{ellingsen09}
S.~\AA.\ Ellingsen, S.~Y.\ Buhmann, and S.~Scheel, Phys.\ Rev.\ A
{\bf 79}, 052903 (2009).
\bibitem{ellingsen10}
S.~\AA.\ Ellingsen, S.~Y.\ Buhmann, and S.~Scheel, Phys.\ Rev.\ Lett.\
{\bf 104}, 223003 (2010).
%SAE: updated reference
\bibitem{ellingsen11}
S.~\AA.~Ellingsen, S.Y.~Buhmann, and S.~Scheel, Phys.\ Rev.\ A {\bf 84}, 060501(R) (2011).
\bibitem{nakajima97}
T.~Nakajima, P.~Lambropoulos and H.~Walther, Phys.\ Rev.\ A
\textbf{56}, 5100 (1997).
\bibitem{wu00}
S.-T.\ Wu and C.~Eberlein, Proc.\ R.\ Soc.\ Lond.\ Ser.\ A
\textbf{456}, 1931 (2000).
\bibitem{gorza06}
M.-P.\ Gorza and M.~Ducloy, Eur.\ Phys.\ J.\ D {\bf 40}, 343 (2006).
\bibitem{antezza08}
M.~Antezza, L.~P.\ Pitaevskii, S.~Stringari, and V.~B.\ Svetovoy,
Phys.\ Rev.\ A\ {\bf 77}, 022901 (2008).
\bibitem{sherkunov09}
Y.~Sherkunov, Phys.\ Rev.\ A {\bf 79}, 032101 (2009).
\bibitem{buhmann07}
S.~Y.\ Buhmann and D.-G.\ Welsch, Prog.\ Quantum Electron.\ {\bf 31},
51 (2007).
\bibitem{scheel08}
S.~Scheel and S.~Y.\ Buhmann, Acta Phys.\ Slov.\ {\bf 58}, 675
(2008).
\bibitem{ellingsen10Q} 
S.\ {\AA}.\ Ellingsen, Y.\ Sherkunov, S.Y.\ Buhmann, and S.\ Scheel,
in \textit{Proceedings of the Ninth Conference on Quantum Field Theory
Under the Influence of External Conditions (QFEXT09)} edited by M.\
Bordag and K.~A.\ Milton (World Scientific, 2010), p.\ 168, Preprint:
\texttt{quant-ph/0910.5608}.
\bibitem{weber10}
A.~Weber and H.~Gies, Phys.\ Rev.\ Lett.\ \textbf{105}, 040403 (2010).
\bibitem{ferrell80}
T.~L.\ Ferrell and R.~H.\ Ritchie, Phys.\ Rev.\ A {\bf 21}, 1305
(1980).
\bibitem{marvin82}
A.~M.\ Marvin and F.~Toigo, Phys.\ Rev.\ A {\bf 25}, 782 (1982); {\bf
25}, 803 (1982).
\bibitem{girard89}
C.~Girard, S.~Maghezzi, and F.~Hache, J.~Chem.\ Phys.\ {\bf 91}, 5509
(1989).
\bibitem{jhe95}
W.~Jhe and J.~W.\ Kim, Phys.\ Rev.\ A {\bf 51}, 1150 (1995).
\bibitem{klimov96}
V.~V.\ Klimov, M.~Ducloy, and V.S.\ Letokhov, J.~Mod.\ Opt.\ {\bf 43},
2251 (1996).
\bibitem{buhmann04}
S.~Y.\ Buhmann, H.~T.\ Dung, and D.-G.\ Welsch, J.~Opt.\ B: Quantum
Semiclass.\ Opt.\ {\bf 6}, S127 (2004); Erratum: J.~Phys.\ B: At.\
Mol.\ Opt.\ Phys.\ {\bf 39}, 3145 (2006).
\bibitem{taddei10}
M.~M.\ Taddei, T.~N.~C.\ Mendes and C.~Farina, Eur.\ J.\ Phys.\ {\bf
31}, 89 (2010).
\bibitem{sambale10}
A.~Sambale, S.~Y.\ Buhmann, and S.~Scheel, Phys.\ Rev.\ A {\bf 81},
012509 (2010).
\bibitem{babikker76}
M.\ Babiker and G.\ Barton, J.\ Phys.\ A:\ Math.\ Gen.\ \textbf{9},
129 (1976).
\bibitem{brevik06}
I.~Brevik, S.~A.\ Ellingsen and K.~A.\ Milton, New J.\ Phys.\ {\bf 8},
236 (2006).
\bibitem{buhmann10}
S.~Y.\ Buhmann, S.~Scheel and J.~Babington, Phys.\ Rev.\ Lett.\
\textbf{104}, 070404 (2010).
\bibitem{abramowitz}
M.~Abramowitz and I.~A.\ Stegun, {\it Handbook of Mathematical
Functions} (Dover, New York, 1964).
\end{thebibliography}
\end{document}